\documentclass[twocolumn,showpacs,preprintnumbers,amsmath,amssymb]{revtex4}
\usepackage{graphicx}% Include figure files
\usepackage{dcolumn}% Align table columns on decimal point
\usepackage{bm}% bold math
\begin{document}
\title{Magnetic reconnection during collisionless, stressed,  X-point collapse  using Particle-in-Cell
simulation} 
\author{D. Tsiklauri and T. Haruki}
\affiliation{Institute for Materials Research, University of Salford, Manchester, M5 4WT, Great Britain}
\date{\today}
\begin{abstract}
Magnetic reconnection during collisionless, stressed, X-point collapse was studied 
using kinetic, 2.5D, fully electromagnetic, relativistic Particle-in-Cell numerical code. 
Two cases of weakly and strongly stressed X-point collapse were considered.
Here descriptors weakly and strongly refer to 20 \% and 124 \% unidirectional spatial compression of the X-point, respectively. 
In the weakly stressed case,
the reconnection rate, defined as the out-of-plane electric field in the X-point (the magnetic null) 
normalised by the product of external magnetic field and Alfv\'en speeds, peaks at 0.11, with its 
average over 1.25 Alfv\'en times being 0.04. During the peak of the reconnection, electron inflow 
into the current sheet is mostly concentrated along the separatrices until they deflect from 
the current sheet on the scale of electron skin depth, with the electron outflow speeds being 
of the order of the external Alfv\'en speed. Ion inflow starts to deflect from the current 
sheet on the ion skin depth scale with the outflow speeds about four times smaller than that of electrons.
Electron energy distribution in the current sheet, at the high-energy end of the spectrum, 
shows a power law distribution with the index varying in time, attaining a maximal value of $-4.1$ 
at the final simulation time step (1.25 Alfv\'en times). 
In the strongly stressed case, magnetic 
reconnection peak occurs 3.4 times faster and is more efficient.
The peak reconnection rate now attains value 2.5, with the average reconnection rate over 
1.25 Alfv\'en times being 0.5. Plasma inflow into the current sheet is perpendicular to it, 
with the electron outflow seeds reaching 1.4 Alfv\'en external Mach number 
and ions again being about four times slower than electrons. 
The power law energy spectrum for the 
electrons in the current sheet attains now a steeper index of $-5.5$, a value close to the
ones observed near X-type region in the Earth's magneto-tail. 
Within about one Alfv\'en time,  2\%  and 20\% of the initial magnetic energy is converted 
into heat and accelerated particle energy in the case of weak and strong stress,
respectively. 
In the both cases, during the peak of the reconnection, the quadruple out-of-plane magnetic field 
is generated, hinting possibly to the Hall regime of the reconnection. 
These results strongly suggest the importance of the 
collisionless, stressed X-point collapse as an efficient mechanism of converting 
magnetic energy into heat and super-thermal particle energy.
\end{abstract}


\maketitle

\section{Motivation of the study}

Magnetic reconnection is an important physical process, which
serves as one of the possible ways of converting energy stored
in the magnetic field into heat and non-thermal, accelerated,
motion of plasma particles. This process operates virtually in
all extra-galactic, stellar, solar, space and laboratory plasmas
with varied degree of importance. 
For example, in solar and stellar flares magnetic
reconnection plays a key role. In addition, it can be
one of the main contributing factors to solar coronal heating
problem amongst other mechanisms such as wave dissipation.
As far as plasma heating is concerned, in the collisional regime,
be it Tokamak plasma or solar corona (which perhaps is better described as
a collisionless medium), Spitzer resistivity is
$\propto T_e^{-3/2}$, where  $T_e$ is the electron
temperature. Thus, in becoming hotter, plasma essentially
starts to behave as a superconductor, i.e. further heating (increase in temperature) 
is impeded due to decrease of the resistive properties.
In this context, a direct conversion of magnetic energy into heat via
reconnection seems rather attractive.
As far as the particle acceleration is concerned, 50-80 \% of the
energy released during solar flares is converted into the energy of accelerated
particles. Thus, knowledge of the details of particle acceleration via magnetic
reconnection (which is deemed as a major mechanism operating solar
and stellar flares) is also important.

The main aspects of the reconnection process can be
classified as whether it is resistive (collisional) or collisionless;
spontaneous or forced;
and/or steady or time-dependent.
Each of these aspects have been studied extensively with varied
level of progress. For example, both steady and time-dependent, {\it resistive} reconnection
processes are
very well studied [see e.g. Ref.\cite{pf00} and references therein], while it is only
recently some progress has been made in the study of {\it collisionless}
reconnection [see e.g. Ref.\cite{b05,bp07} and references therein].
This can be explained by the fact that resistive reconnection is
simpler in its nature. Changing structure (connectivity) of magnetic field lines, which
essentially is the reconnection, requires some mechanism (dissipation) to break the frozen-in
condition. In the case of resistive reconnection it is $\eta \vec j$ 
term. In fact, rate at which reconnection proceeds is given by
an inflow velocity into the dissipation region
$v_{in} \simeq (\delta / \Delta) v_A$, where $\delta$ and $\Delta$ are
width and length of the region, and $v_A$ is Alfv\'en speed.
In the simplest resistive, steady model of Sweet-Parker
$\delta \simeq \eta^{1/2}$ and $\Delta \simeq L$, where $L$ is the
macroscopic system size.
In the collisionless regime however, other terms in the Ohm's law,
such as electron inertia term, Hall term
and electron pressure tensor become far more important than the usual
$\eta \vec j$ term on which resistive reconnection models rely.
Each of these terms has an associated spatial scale with them, 
i.e. scale at which
they start to play a dominant role.
For example, electron inertia term is associated with the electron skin depth
$c/\omega_{pe}$, Hall term is with the ion skin depth $c/\omega_{pi}$, while
electron pressure tensor is with ion Larmor radius [see e.g. p. 87 in 
Ref.\cite{bp07} or p. 42 in Ref.\cite{pf00}]. 

One of the main outcomes of recent research in collisionless reconnection 
can be summarised by GEM \cite{birn2001,pritchett2001} and Newton \cite{birn2005} reconnection challenges.
These considered the same physical system: Harris-type equilibrium with anti-parallel magnetic field, relevant to
geomagnetic tail application, with finite initial magnetic perturbations in the case of GEM challenge and
time-transient inhomogeneous, driven inflow of magnetic flux in the case of Newton challenge.
The novelty of the approach was to use different numerical codes
[MHD, Hall MHD, Hybrid and Particle-in-Cell (PIC)], in order to pin-point essential physical
mechanism that facilitates the reconnection. These works established that
as long as dispersive whistler waves are included (these can only appear if one
allows for the different dynamics for electrons and ions), the rate at which reconnection
proceeds does not change, irrespective of which term breaks the frozen-in condition.

Yet another interesting analytical result corroborated by the numerical simulations is
that in 2D steady reconnection, Petschek reconnection rate functional form
remains the same when the Hall term is included (in the Hall MHD model), but what changes is the length from the X-point to
the start of slow mode shocks \cite{ah06}. In MHD, this length is the half width
of the resistive region, $L_R$, but when Hall term is included the length is
replaced by 	$L_R + c / \omega_{pi}$ \cite{ah06}.
This important result essentially provides an analytical scaling for
the collisionless reconnection (when $\eta$ is zero) for such physical system. 
Thus, future kinetic studies that use PIC simulation need to corroborate it. 
Ref.\cite{ah06}  already tested this scaling law via MHD and Hall MHD simulation.

Importance of the electron inertia in the X-point with finite out-of-plane guide
magnetic field has been investigated in Ref.\cite{ken06}.
Particularly noteworthy result was that when the
normalised collisionless electron skin 
depth [$c/(\omega_{pe}L)$] exceeds the dimensionless 
resistive length scale 
($S^{-1/2}$), where $S$ is the Lindquist number, the energy in
a shear Alfv\'en wave approaching an X-point is rapidly 
transformed into plasma kinetic energy and heat. 
Assuming solar coronal parameters, the normalised electron skin depth
is $10^{-8}$ (assuming $L=10$ Mm and $c/\omega_{pe}= 0.1$ m),
while dimensionless resistive length scale is $3 \times 10^{-7}$ (in the corona typically $S=10^{13}$).
Thus, the proposed in \cite{ken06} may well be effective.

 Previous works on particle acceleration mostly focused on test-particle type
  calculations of the particle trajectories in different 2D \cite{vb97,bv01,zhar04,ham05} or more recently 
  3D \cite{db05,db06,lit06,bd07} magnetic reconnection configurations.
  In such approach feedback on reconnection electromagnetic (EM) fields
  from motion (spatial redistribution) of charged particles
  is ignored. Our approach does not suffer from this drawback, as we use fully electromagnetic, 
  relativistic PIC numerical code in which EM fields are calculated at each step from the
  spatial distribution of the charges.

According to Ref.\cite{pf00}, resistive time-dependent 
reconnection other than well known tearing mode can be
split into two main classes: X-type collapse, which was first considered by
Ref.\cite{d53} (a decade before tearing mode was discovered) and Petschek-type
theory developed by Ref.\cite{s83}. Previous work on X-point collapse
is well described in chapter 7.1 of Ref.\cite{pf00}. Also see more recent work
on the subject \cite{vb05}. A good example of combination of analytical and
numerical work on magnetic reconnection at stressed X-type neutral points can be
found in Ref.\cite{oms93}. Boundary conditions used in Ref.\cite{oms93}, where such
that they did not allow for flux or mass flow through the boundary.
The studies with closed boundaries are physically justified by being isolated systems, 
whereas some of those with open boundaries lead to misleading results: 
for example, in an open system a potential X-point can collapse 
due to inflow of energy from outside.

Initial analytical work on this topic considered 
unbounded self-similar
solutions.  These indicated that $E_z (0,0,t)$, the out-of-plane electric field at the magnetic null, 
which is the measure of reconnection
  rate, tends to infinity as $t \to \infty$ [e.g. Fig.(7.3) from Ref.\cite{pf00}].
  The main outcomes of the previous, stressed X-point collapse 
  in the resistive MHD (in the case of low-resistivity and low-beta) 
  can be summarised as following:

  (i) The X-point collapse is different depending whether initial stressing
  is  weak or strong. In the weak case the average reconnection rate
  scale as $1/\ln(\bar \eta)$, where $\bar \eta =1/S$ 
   is the dimensionless resistivity [$\bar \eta =\eta / (V_{A0} L)=1/S$], 
  while strong case it is independent of $\bar \eta$. 
    
  (ii) There is an issue related to efficiency of the process:
Ref.\cite{mcc96} showed that for fast reconnection to occur
$\beta < \bar \eta^{0.565}$, otherwise pressure in the
current sheet chokes off the collapse process.
For solar coronal conditions, $\bar \eta \simeq 10^{-13}$ this requires
$\beta < 10^{-8}$, which is too low to match the observed values circa $0.01 - 0.001$.
However, this shortcoming seems to be alleviated by inclusion
of non-linear effects, i.e. case of strong perturbations
(strong stressing) \cite{mcc96}. If the compression is 
sufficiently large, magnetic pressure of the imploding
  wave expels trapped gas in the current sheet from its ends, allowing it to thin
  and faster reconnection to occur.

Motivation of the present study is five-fold:

(i) Naturally, different boundary and initial conditions 
produce different scaling of reconnection rate e.g. with
resistivity (in the case of resistive reconnection). To our knowledge
X-type collapse by a uniform stress (compression) along one of the axis,
as described in chapter 2.1 in Ref.\cite{pf00}, has not been
investigated numerically neither in the case of resistive (MHD) 
reconnection, nor in the collisionless regime;

(ii) This type of stress (compression) is likely to occur in the
framework of a flare model (see details below).

(iii) To our knowledge the most of previous collisionless reconnection studies
considered anti-parallel, Harris type magnetic field configuration, which
is more relevant to Earth magneto-tail application. For solar and stellar
flares X-type configuration is more relevant.

(iv) Some of the models of coronal mass ejections \cite{fp95} which
  use motion of photospheric footpoints as a driver, end up
  with situation physically similar to stressed X-point collapse (strictly
  speaking Y-points occur there).
  
(v) We aimed to investigate properties of the accelerated particles in the
current using self-consistent electromagnetic fields, as a further extension of the
test-particle type approach.

\section{Simulation model}

\subsection{Stressed X-point reconnection model}

  \begin{figure}
    \resizebox{\hsize}{!}{\includegraphics{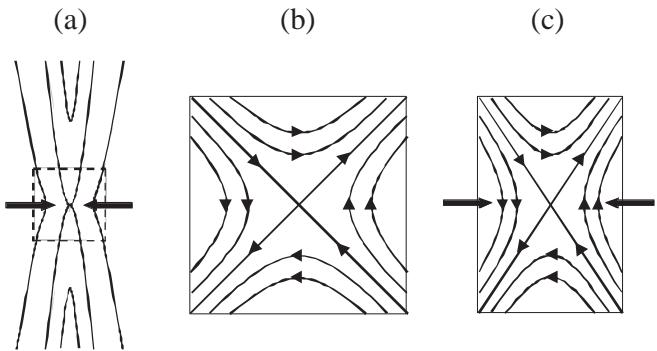}}
     \caption{ \label{Fig:model}Magnetic field line geometry.
             (a) A portion of a standard Solar flare model 
	     with added compressions on sides. 
	     The bottom part mimics the footpoints of a coronal loop,
	     with an X-point on top (inside the dashed box).
             Two arrows indicate compression direction. 
             A dashed box is the region which 
	     our study tries to mimic by uniform stress in one direction 
	     [see (c)].
	     (b) Magnetic field configuration 
             for stress parameter $\alpha = 1$, which is stable (no X-point collapse
	     occurs). 
             (c) Magnetic field configuration 
             for stress parameter $\alpha > 1$.
             Such X-point collapses under high magnetic pressure/tension forces.}
      \end{figure}
  
  Fig.~\ref{Fig:model} represents magnetic field line geometry considered 
  in this paper. In order to study stressed magnetic reconnection, 
  we consider a magnetic X-point collapse which may naturally 
  occur during solar flares [see sketch in Fig.~\ref{Fig:model}(a)]. 
  Such stressed X-point may occur e.g. if 
  photospheric footpoints of the coronal loops move
  towards each other (e.g. pushed by convective motions)
  or some compression from the sides in the corona
  forces them to do so. We study  dynamics of such stressed X-point 
  by means of kinetic, 2.5D, fully electromagnetic, 
  relativistic Particle-in-Cell numerical code.
  We focus attention on the local region inside the dashed box which is
  is the region that our study tries to mimic by 
  uniform stress in one direction   
  [see Fig.~\ref{Fig:model}(c)].
  If there is no stress from the sides the considered magnetic
  configuration is stable [Fig.~\ref{Fig:model}(b)].
  Interestingly, Ref.\cite{sakai_tanaka2007}   
  investigated what happens when a fast magnetosonic shock wave 
  associated with a coronal mass ejection collides 
  obliquely with a coronal streamer with a stable current sheet.
  Their set up is somewhat analogous to ours, but much more violent,
  as we only consider X-point compression, while they blasted it with 
  a Alfv\'en Mach 6 shock.
    
   The initial magnetic field configurations used are
    \begin{equation}
    B_x = \frac{B_0}{L} y,
    \;\;\;    B_y = \frac{B_0}{L} \alpha^2 x, \;\;\; 
     B_z = 0,
    \end{equation}
  where $B_0$ is magnetic field intensity at the distance $L$ from the X-point 
  for $\alpha = 1.0$, $L$ is the global external length-scale of reconnection, 
  and $\alpha$ is the stress parameter [see chapter 2.1 in Ref.\cite{pf00}].
  In addition, the uniform current is imposed at $t=0$ in the $z$-direction,
  \begin{equation}
    j_z = \frac{B_0}{\mu_0 L} (\alpha^2 - 1).
   \end{equation}
  This current appears because of the stress imposed on magnetic field
  lines. 
   
  \subsection{PIC simulation code}
  
  The simulation code used here is 2.5D relativistic and fully electromagnetic 
  PIC code, modified from the 3D TRISTAN code 
  \cite{buneman1993}. 
  In this code, both the electron and ion dynamics are described as particles.
  Equation of motion for each particle is 
  solved using self consistent electromagnetic fields,
    \begin{equation}
    \frac{d \vec{v}_{si}}{d t} = \frac{q_s}{m_s}(\vec{E} 
    + \vec{v}_{si} \times \vec{B}),
    \label{eq_motion}
  \end{equation}
    \begin{equation}
    \frac{d \vec{r}_{si}}{d t} = \vec{v}_{si},
    \label{eq_position}
  \end{equation}
    \begin{equation}
    \frac{\partial \vec{E}}{\partial t} = c^2 \nabla \times \vec{B} 
    - \frac{1}{\epsilon_0} \Sigma_s \vec{j}_s,
    \label{eq_ampare}
  \end{equation}
    \begin{equation}
    \frac{\partial \vec{B}}{\partial t} = - \nabla \times \vec{E},
    \label{eq_faraday}
  \end{equation}
  where $\vec{E}$, $\vec{B}$, $\vec{j}_s$, $\vec{v}_{si}$ and $\vec{r}_{si}$ 
  are electric and magnetic fields, current density, particle velocity 
  and position, respectively. 
  The subscript $s$ represents species of plasma, that is $s = e$ for electron 
  and $s = i$ for ion.
  The subscript $i$ indicates $i$-th particle index.
  The other quantities,  $q_s$, $m_s$, $c$, $\epsilon_0$, are 
   charge and mass of a plasma particle, speed of light, 
  vacuum permittivity, respectively.
  In addition to above Eqs.~(\ref{eq_motion})--(\ref{eq_faraday}), 
  $\nabla \cdot \vec{B} = 0$ and $\nabla \cdot \vec{E} = \rho_e / \epsilon_0$ 
  must be satisfied initially ($\rho_e$ is charge density, which is taken as 0). 
 The latter two conditions are automatically satisfied at all times
 due to the nature of the numerical scheme used \cite{buneman1993}.
 Electromagnetic fields which satisfy Eqs.~(\ref{eq_ampare})--(\ref{eq_faraday})
 update particle velocity through the equation of motion 
 Eq.~(\ref{eq_motion}), which in turn updates particle positions
 Eq.~(\ref{eq_position}). Repeating this
 procedure many times mimics plasma particle dynamics
 in self-consistent electromagnetic fields.
  
  \begin{figure}
     \resizebox{\hsize}{!}{\includegraphics{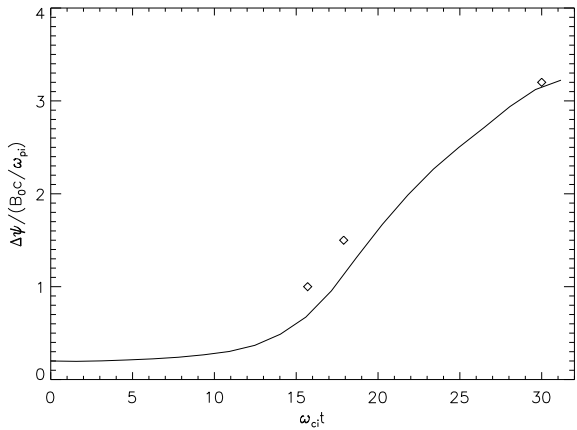}}
     \caption{\label{Fig:gem} Time evolution of the reconnected magnetic flux difference.
             The horizontal axis is time normalised by 
             the ion cyclotron frequency, $\omega_{ci}$.
             The solid curve is the magnetic flux difference, $\Delta \psi$, 
             between the O and X lines, which is normalised 
             by $B_0 c/\omega_{pi}$, using our PIC code. Open squares correspond to
	     published GEM challenge values.}
       \end{figure}

  We started by testing our code, reproducing published 
  GEM challenge results \cite{birn2001, pritchett2001}.
  Fig.~\ref{Fig:gem} shows time evolution of
  the reconnected magnetic flux
  difference between and O and X-lines.
  Open squares correspond to the published GEM challenge values.
  We gather from this graph that match between our simulation results
  and that of GEM challenge \cite{birn2001, pritchett2001} is   good.
  We could also reproduce other figures from Refs.\cite{birn2001, pritchett2001},
  thus we were reasonably confident to start a new 
  X-point collapse simulation, which is the subject of the present paper.
  
  The length of the system in two dimensions is
   $L_x = L_y = 400 \Delta$ (this is excluding, so called,
  ghost cells needed for updating the boundary conditions),
  where $\Delta = 1$ is the simulation grid size corresponding to electron 
  Debye length, $\lambda_D = v_{te} / \omega_{pe} = 1 \Delta$ ($v_{te}$ is
   electron thermal velocity and $\omega_{pe}$ is electron plasma 
  frequency). The global external length-scale of reconnection is set 
  $L = 200 \Delta$. Number density is fixed at $n_0 = 100$ electron-ion 
  pairs per cell. Both electrons and ions are uniformly distributed 
  throughout the system, hence total number is $1.6$ million pairs.
  
  Zero-gradient boundary conditions are imposed 
  both on the electric and magnetic fields 
  in  $x$- and $y$-directions. Also, tangential component of electric field 
  was forced to zero, while normal component of magnetic field was kept
  constant, both at the boundary.
  This ensures that there is no magnetic flux through the simulation boundary,
  i.e. the system is isolated and 
  $
  -\int_{-L}^{L} B_y(x',-L,t)dx'
  -\int_{-L}^{L} B_y(x',L,t)dx'+
    \int_{-L}^{L} B_x(-L,y',t)dy'+
  \int_{-L}^{L} B_x(L,y',t)dy'=0,
  $
  at every time step. 
  The latter sum is the magnetic flux on the boundary.
  Reflecting boundary conditions are imposed on particles 
  in both the $x$- and $y$-directions. The latter ensures there is no
  mass flow across the boundary.
  
  The simulation time step is $\omega_{pe} \Delta t = 0.05$. 
  Ion to electron mass ratio is $m_i / m_e = 100$.
  Electron thermal velocity to speed of light ratio is $v_{te} / c = 0.1$.
  Electron and ion skin depths are $c / \omega_{pe} = 10 \Delta$ and 
  $c / \omega_{pi} = 100 \Delta$, respectively. 
 Electron cyclotron frequency to plasma frequency ratio is 
  $\omega_{ce} / \omega_{pe} = 1.0$ for magnetic field intensity, $B = B_0$.
  The latter ratio is indeed close to unity in the solar corona, while it is much
  bigger than unity in the Earth magnetosphere.
  Electron and ion Larmor radii are 
  $v_{te} / \omega_{ce} = 1 \Delta$ and $v_{ti} / \omega_{ci} = 10 \Delta$, 
  where $v_{ti}$ is the ion thermal velocity. 
  The temperatures of ions and electrons are the same, $T_e = T_i$.
 At the boundary the plasma $\beta=0.02$ and 
 Alfv\'en velocity, $V_{A0} / c = 0.1$.
 Naturally these vary across the simulation box as 
 the background magnetic field is a function of $x$ and $y$.
 
 In what follows, for the visualisation purposes, all 
 spatial coordinates will be normalised by electron skin depth
  $c/ \omega_{pe}$, while time is normalised by the inverse of 
  plasma electron frequency 
  $\omega_{pe}^{-1}$. 
 
 \section{Simulation results}

  We investigated three X-point collapse cases 
  with different stress parameters, $\alpha = 1.00$ (stable), 
  $1.20$ (weakly stressed) and $2.24$ (strongly stressed). 
  For $\alpha = 1.00$, we confirmed that the system is stable 
  at least for $t \leq 500$,  and no magnetic reconnection takes place.
  Ref.\cite{mcc96} has shown that when perturbations (exerted stress on an X-point)
  are small $\varepsilon \sim (1-\alpha^2) < \bar \eta$ then 
  the (average) reconnection rate scales as $1/\ln(\bar \eta)$, while
  if they are large $\varepsilon = (1-\alpha^2) > \bar \eta$ then 
  the reconnection rate is independent of $\bar \eta$.
  This difference in behaviour points to a different relative importance
of the  physical mechanisms in action. 
The two different (weakly and strongly compressed) cases studied in this
paper  attempt the same approach as in Ref.\cite{mcc96}.
However, one should realise that we use vastly different
physical description. Ref.\cite{mcc96} uses
resistive MHD, while our PIC (fully kinetic) approach is
collisionless, and, in turn, in our case $\bar \eta =0$.
However, it has been observed before, that often scattering of
plasma particles from the magnetic field lines, plays
effective role of collisions. E.g. Refs.\cite{tss05a,tss05b}
have shown that the Alfv\'en wave dissipation in the collisionless
case (using the same PIC, kinetic code) follows  scaling law which
is the same as in resistive MHD case \cite{hp83}.
At the same time, this is not to say that we are confident that the
effective particle scattering off the magnetic fields, 
which mimics resistive (collisional) effects,
{\it is} the mechanism that breaks down the frozen-in condition here. 
The issue of 
which term in the
generalised Ohm's law is responsible for the reconnection  
in the collisionless
stressed X-point collapse will be studied
separately.

\subsection{Weakly stressed X-point case ($\alpha = 1.20$)}

It should be noted that $\alpha$, the measure of stress, is essentially
   the aspect ratio of the compression, i.e. since the limiting magnetic field lines
   are given by $y=\alpha x$, $\alpha=1.2$ means 1.2-1.0=0.2 i.e. 20\% compression
   of the X-point in $x$-direction. Similarly, $\alpha=2.24$ case represents
   124\% compression, i.e. 1.24 times stressed.  
   
   Also, because change in $\alpha$ means change of strength
   of the magnetic field on the boundary,
   the external Alfv\'en speed also changes.
   E.g. we fixed Alfv\'en velocity, $V_{A0} / c = 0.1$ for $\alpha=1$,
   but for $\alpha>1$, $V_{A0} / c =\sqrt{(1+\alpha^4)/2} \times 0.1$.
   This for $\alpha = 1.20$, $V_{A0} / c = 0.124$, but for 
   $\alpha = 2.24$, $V_{A0} / c = 0.36$.

  \begin{figure}
     \resizebox{\hsize}{!}{\includegraphics{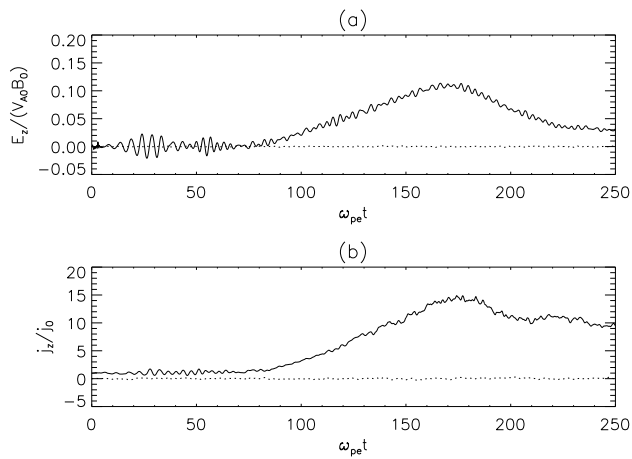}}
      \caption{\label{Fig:timed_ezjz_a120}(a) Time evolution of 
    the reconnection rate,  
defined as the out-of-plane electric field in the X-point, $E_z(0,0,t)$ normalised by 
the product of external magnetic field and Alfv\'en speeds (both at the boundary for $\alpha = 1.00$)
             for $\alpha = 1.20$ (solid line) and $\alpha = 1.00$ (dotted line). 
             (b) Time evolution of total current density, $j_z$, 
             in the X-point for $\alpha = 1.20$.
             Note that final simulation time 
	     $t=250$ corresponds to 
	     $t/\tau_A=1.25$, with the latter being the Alfv\'en time.}
      \end{figure}
  
  Fig.~\ref{Fig:timed_ezjz_a120}(a) shows 
  time evolution of the reconnection rate,  
defined as the out-of-plane electric field in the X-point, $E_z(0,0,t)$ normalised by 
the product of external magnetic field and Alfv\'en speeds (both at the boundary), i.e.
by $E_0=V_{A0}B_0$,
  for $\alpha = 1.20$ (solid line) and $\alpha = 1.00$ (dotted line). 
 Because usually PIC simulation suffers for large (thermal) noise especially
towards high energy ends of the particle distribution function,
a boxcar average scheme 
with width $150$ mesh points was applied 
for smoothing the line-data.
Mind that this does not alter absolute value of the
peak at $t=170$.
Our normalisation of the electric field is such that
effectively Fig.~\ref{Fig:timed_ezjz_a120}(a)
shows the Alfv\'en Mach number of the
inflowing plasma, $M_A$, i.e. $E/E_0=v/V_{A0}=M_A$.
This follows from a typical estimate $E \simeq v B_0 \simeq M_A V_{A0} B_0$ [e.g. Ref.\cite{bv01}].
We gather from this figure that  for the case of $\alpha=1$
no $E_z$ is generated, as such configuration is stable and
no X-point collapse or reconnection occurs (dotted line).
On contrary, for the case of  $\alpha=1.2$ we see that the 
reconnection rate peaks at about 0.11 at $t=170$ which is the same as $t/\tau_A=0.85$.
Here $\tau_A$ is the Alfv\'en time -- the distance from the outer boundary
to the X-point ($20$) divided by the Alfv\'en speed at the
boundary ($V_{A0}=0.1c$).
We also calculated the average reconnection rate based on Fig.~\ref{Fig:timed_ezjz_a120}(a).
It is defined as  
\begin{equation}
E_{av}=\frac{1}{t_f}\int_0^{t_f}\frac{E_z(0,0,t)}{E_0} dt, 
\end{equation}
where $t_f=250$  
is the final simulation time. This yields $E_{av}=0.04$, a relatively small value.

Fig.~\ref{Fig:timed_ezjz_a120}(b) shows time evolution of 
  total current density, $j_z$, in the X-point for $\alpha = 1.20$ (solid line)
  and $\alpha=1.0$ (dotted line).
 The total current density is normalised by the initial value, 
$j_0 = n_0 e v_{d0}$,  $v_{d0}$ is 
the drift velocity.
As expected $\alpha=1$ case stays current free throughout the simulation,
while  $\alpha=1.2$ produces a peak current of $j_z/j_0=15$ and then subsequently decays off.

It should be noted that we have performed one additional numerical run with
$\alpha=1.2$, but without imposing initial $j_0$ current prescribed by Eq.(2).
This is because e.g. one might expect that in a X-point above the arcade
of loops in the solar corona, only compression (stress) of the magnetic
field from the two sides is likely. Such perturbation violates the
equations at $t=0$, this results in a large spike of magnitude 0.22 at
$t=5$ in the equivalent version of Fig.~\ref{Fig:timed_ezjz_a120}(a) (not presented here).
Otherwise, for $t > 5$ there was no noticeable difference
between the cases with and without imposing the initial current. Thus, in what follows
we only discuss cases with the current imposed at $t=0$.
    \begin{figure}
    \resizebox{\hsize}{!}{\includegraphics{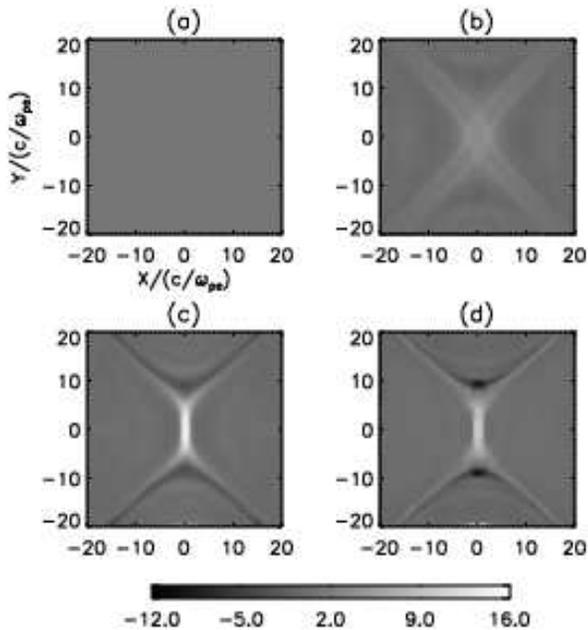}}
     \caption{\label{Fig:jz_a120}Time evolution  of the spatial distribution of total current density, 
             $j_z$, in the $X$-$Y$ plane at (a) $t = 0$, 
             (b) $100$, (c) $170$ and (d) $250$ for $\alpha = 1.20$.
             The total current density is normalised by the initial value, 
             $j_0 = n_0 e v_{d0}$.}
     \end{figure}
   
    Fig.~\ref{Fig:jz_a120} shows time evolution of spatial distribution of  
  total current density, $j_z$, in the $x$-$y$ plane at (a) $t = 0$, 
  (b) $100$, (c) $170$ and (d) $250$ for $\alpha = 1.20$.
  The current flows uniformly in the $z$-direction Fig.~\ref{Fig:jz_a120}(a), 
  with drift velocity is $v_{d0} / c = 0.022$.
  We assumed that half of the drift velocity comes from electrons and another half
  from ions, i.e. $v_{de0}=0.5v_{d0}$ and $v_{di0}=0.5v_{d0}$.
  Fig.~\ref{Fig:jz_a120}(b) shows that the current begins to focus in the X-point 
  because of high magnetic pressure/tension. 
  The current peaks in the current sheet attaining $j_z / j_0 = 15$ 
  at $t = 170$, as seen in Fig.~\ref{Fig:jz_a120}(c).
  In the later phases, the current sheet tends to decay [Fig.~\ref{Fig:jz_a120}(d)].
  This figure is useful for visualising the spatial dimensions 
  of the current sheet. In turn, this enables to make
  the following useful estimate: From Fig.~\ref{Fig:jz_a120}(c) (peak current time
  snapshot) we gather that the width of the current sheet (in horizontal direction) is about 
  electron skin-depth $\delta \simeq 1.4$, while its length 
  (in vertical direction) is about $\Delta \simeq 10$.
  To be precise, we actually measured $\delta$ and $\Delta$ by more accurate
  means: we looked at line plots of $j_z(x,0)$ and $j_z(0,y)$ respectively, at
  $t=170$, and measured appropriate half-width of the $j_z(x,0)$ and $j_z(0,y)$  
  peak (in both $x$ and $y$-directions).  
  The ratio of the two gives the inflow  Alfv\'en Mach number $M_A \simeq \delta / \Delta \simeq 0.14$.
  This is very close to the peak value of $E/E_0=v/V_{A0}=M_A=0.11$ [from Fig.~\ref{Fig:timed_ezjz_a120}(a)]. 
  We thus measured
  the peak reconnection rate by
  two {\it independent} means directly from the simulation [using Fig.~\ref{Fig:timed_ezjz_a120}(a)] and using
  the steady reconnection formula $M_A \simeq \delta / \Delta$  [e.g. Eq.(3.1) from Ref.\cite{bp07}].
  The close match points to the fact that nearly steady reconnection regime is achieved
  in the vicinity of the peak at $t=170$.

  \begin{figure}
    \resizebox{\hsize}{!}{\includegraphics{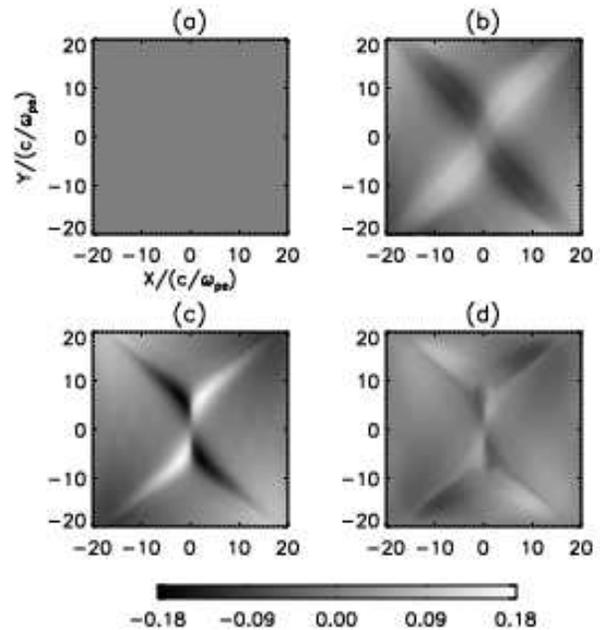}}
        \caption{\label{Fig:bz_a120} Time evolution of the spatial distribution of 
    the out-of-plane magnetic field,  $B_z$, at (a) $t = 0$, 
             (b) $100$, (c) $170$ and (d) $250$ for $\alpha = 1.20$.
             The magnetic field intensity is normalised by the initial value, 
             $B_0$.}
      \end{figure}
  
  Fig.~\ref{Fig:bz_a120} shows the time evolution  of spatial distribution of the
  out-of-plane magnetic field, $B_z$, at (a) $t = 0$, 
  (b) $100$, (c) $170$ and (d) $250$ for $\alpha = 1.20$.
  Time of each panel in Fig.~\ref{Fig:bz_a120} corresponds to those of 
  Fig.~\ref{Fig:jz_a120}, respectively.
  As can be seen in  Fig.~\ref{Fig:bz_a120}(a) initially there is
  no out-of-plane magnetic field present.
  Then, Fig.~\ref{Fig:bz_a120}(b) shows that the quadruple magnetic field 
  structure appears, which is the most pronounced at  
  $t=170$ [Fig.~\ref{Fig:bz_a120}(c)].
  This structure is a well known signature of a magnetic reconnection
  in the Hall MHD regime. Ref.\cite{uk06} showed that as long as one allows for
  different electron and ion dynamics (two-fluid description) i.e. when
  Hall term is non-zero such quadruple out-of-plane magnetic field is
  generated.
  $B_z$  attains  a value of $|B_z| / B_0 = 0.18$ at $t=170$ and then subsequently decays, Fig.~\ref{Fig:bz_a120}(d),
  (as well as the current sheet).
  
  \begin{figure}
    \resizebox{\hsize}{!}{\includegraphics{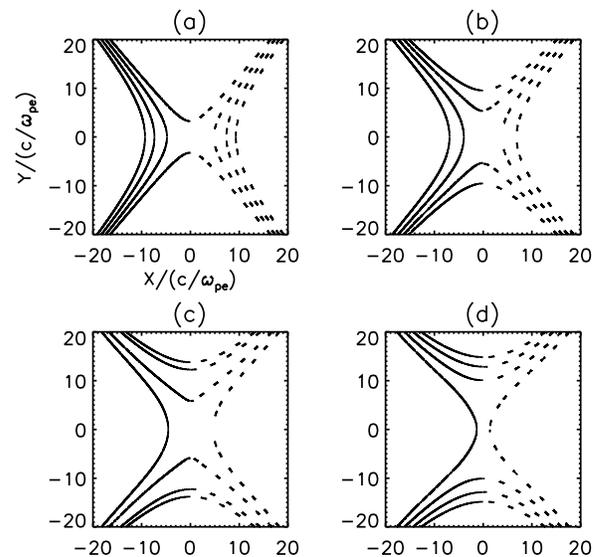}}
        \caption{ \label{Fig:bl_a120} Time evolution of the magnetic field lines 
             in the $x$-$y$ plane at (a) $t = 0$, (b) $100$, 
             (c) $170$ and (d) $250$ for $\alpha = 1.20$.
             The several lines are plotted with different intensity, 
             $|B| / B_0 = 1.55$, $1.60$, $1.65$ and $1.70$ on the boundaries.
             The left and right field lines in the simulation box 
             are distinguished 
             with solid and dotted line styles, respectively, 
             to visualise magnetic field reconnection clearly.}
     \end{figure}
  
  In Fig.~\ref{Fig:bl_a120} we show dynamics of individual magnetic field lines.
  We tried to trace dynamics of several magnetic field lines
  in order to visualise the reconnection process.
  We can clearly see that magnetic field lines come towards
  each other in $x$-direction, reconnect, and move apart in $y$-direction.

  \begin{figure}
   \resizebox{\hsize}{!}{\includegraphics{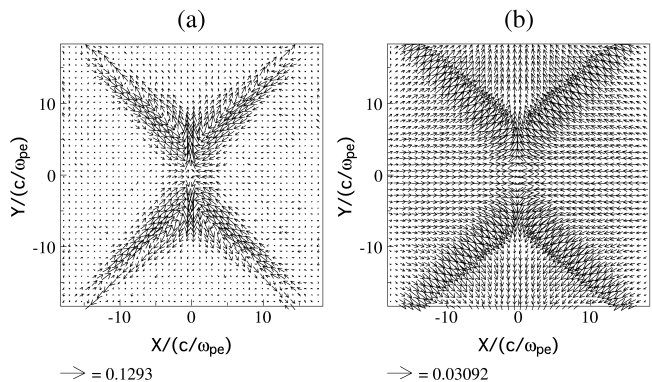}}
   \caption{ \label{Fig:fl_a120}Electron (a) and ion (b) flow pattern around X-point 
    at $t=170$ for $\alpha = 1.20$. Note that here $V_{A0} / c = 0.124$.}
     \end{figure}
  
  Fig.~\ref{Fig:fl_a120} shows electron (a) and ion (b) flows
  at the peak time of the reconnection. It is rather instructive
  to see that this figure in effect corroborates
  the sketch from Ref.\cite{bp07}, [see their Fig.~(3.1)].
  In particular it shows that the electron
  and ion flow are clearly separated on the two different spatial 
  scales -- electron skin-depth and ion skin-depth, note that 
  since here the mass ratio is 100, $c/\omega_{pi}=10 c/\omega_{pe}$).
  The noticeable difference is caused because they considered initial background 
  magnetic field of Harris-type (anti-parallel), while we study X-point.
  Naturally, in our case 
electron inflow into the current sheet is mostly concentrated along
the separatrices until they deflect from the current sheet on the 
scale of electron skin depth, with the electron outflow speeds being of the order of 
the external Alfv\'en speed $0.13c$. 
Ion inflow starts to deflect from the current sheet on the ion skin depth scale with
the outflow speeds about four times smaller ($0.03c$) than that of electrons.
As argued by Ref.\cite{shay01} it is difference in these two flows which
generates the observed quadruple out-of-plane magnetic field. 
  
  \begin{figure}
    \resizebox{\hsize}{!}{\includegraphics{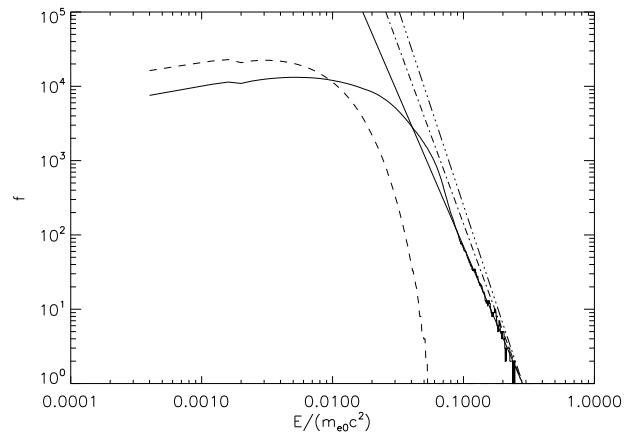}}
    \caption{\label{Fig:ed_a120} The local electron energy spectrum (distribution function) 
             near the current sheet at $t = 0$ (dashed curve) 
             and $t = 250$ (solid curve) for $\alpha = 1.20$. 
             Data is produced using region 
             near the maximum current sheet generation 
             ($-2 \leq x \leq 2, 
               -8 \leq y \leq 8$).
             The vertical axis shows
             the number of electrons with a particular energy. The latter is
	     shown on the horizontal axis (normalised by 
             electron rest energy, $m_{e0} c^2$).
	     Straight solid line is our best fit, while dash-dotted and dash-triple-dotted lines
	     are taken from the observations (see text).}
      \end{figure}
  
  Fig.~\ref{Fig:ed_a120} shows the local electron energy distribution function 
  near the current sheet at $t = 0$ (dashed curve) 
  and $t = 250$ (solid curve) for $\alpha = 1.20$.
  We performed numerical fit to the high energy part of the distribution function
  ($E > 0.08 m_{e0} c^2=41$ keV) and found that the electron distribution has
  a power law form i.e. in particular the best fit is provided by $f=dN/dE \propto E^{-4.1}$
  (straight solid line). In general, the direct observational evidence for the 
  relation between reconnection and  particle acceleration is hard to obtain because
  acceleration length scale is of the order of ion skin-depth.
  However, recently it became possible to directly measure
  electron energy spectrum in the vicinity of X-type region in the Earth's magneto-tail \cite{oe02}.
  They found that from about 2 keV to about 200 keV power law index was between -4.8 and -5.3
  (changing in the course of time of the observation). It is interesting to note that 
  both our power law index (-4.1) and the energy ($0.2 \times m_{e0} c^2 \simeq 0.2 \times 511 \simeq 100$ keV)
  at the high end of the spectrum are close to the observed values by Ref.\cite{oe02}.
  Note that since the parameters of our simulation are
  for the solar corona ($\omega_{ce} / \omega_{pe} = 1.0$), and are not even totally realistic for that 
  (e.g. to simplify numerical
  simulation our $m_i/m_e$, $v_{te} / c$, and $V_{A0} / c$ ratios are not entirely realistic),
  the absolute values of our simulation quoted in keV should be taken with caution (when comparing them to Earth's magneto-tail
  results). Yet we are confident in the correctness of the obtained power law indexes.
  Also, noteworthy fact is that the whistler wave turbulence, based on 
  Fokker-Planck equation for the electron distribution function, 
  subject to a zero-flux boundary condition (same as ours by a chance),
  is producing \cite{ms98} similar power law energy spectrum. Recall that it is
  standing  whistler waves are thought to be mediating reconnection in the Hall regime.

\subsection{Strongly stressed case ($\alpha = 2.24$)}
  
  \begin{figure}
   \resizebox{\hsize}{!}{\includegraphics{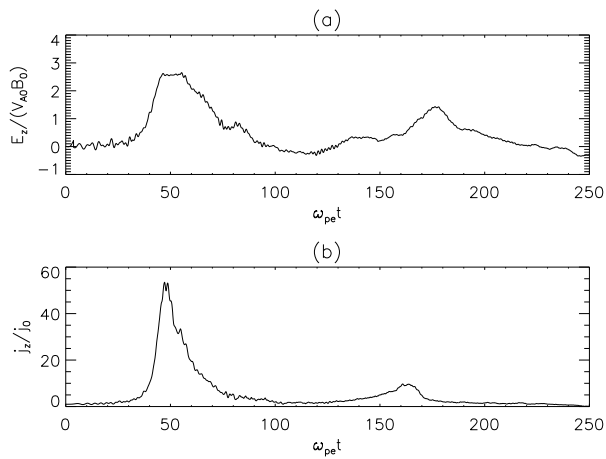}}
      \caption{\label{Fig:timed_ezjz_a224} The same as in Fig.~\ref{Fig:timed_ezjz_a120} 
    but for $\alpha=2.24$. }
      \end{figure}
      
   In this subsection we present results of numerical simulation for the
   case of $\alpha=2.24$, which is regarded as a case of strongly stressed X-point.
     
  Fig.~\ref{Fig:timed_ezjz_a224} shows the numerical simulation results
  as in Fig.~\ref{Fig:timed_ezjz_a120} but for $\alpha=2.24$.
  To avoid repetition we omitted $\alpha=1$ case [dotted line in Fig.~\ref{Fig:timed_ezjz_a120}].
  We gather from panel (a) that now reconnection rate attains value of 2.5 at 
  $t=45$ ($0.225\tau_A$). In the strongly stressed case,
  in differ the weakly stressed one, $E_z$ rebounds and an oscillation is established.
  We calculated average reconnection rate based on Fig.~\ref{Fig:timed_ezjz_a224}(a)
  using definition from Eq.(7). The result is $E_{av}=0.5$, a value $\simeq 10$ times
  bigger than in the weakly stressed case, indicative of more efficient reconnection.
   Also, much stronger current density, $j_z/j_0=55$, is generated compared to the weak case [panel (b)] and
  peak of the reconnection occurs $170 /45 =3.8$ times earlier. 
  Similar  oscillations of current and electric field were used as a
   mechanism for interpreting the peculiar hard x-ray ($> 25$ keV) solar flare, which is believed to be
produced by a non-thermal electron beam \cite{nakariakov2006, ofman_sui2006}.
   
  \begin{figure}
   \resizebox{\hsize}{!}{\includegraphics{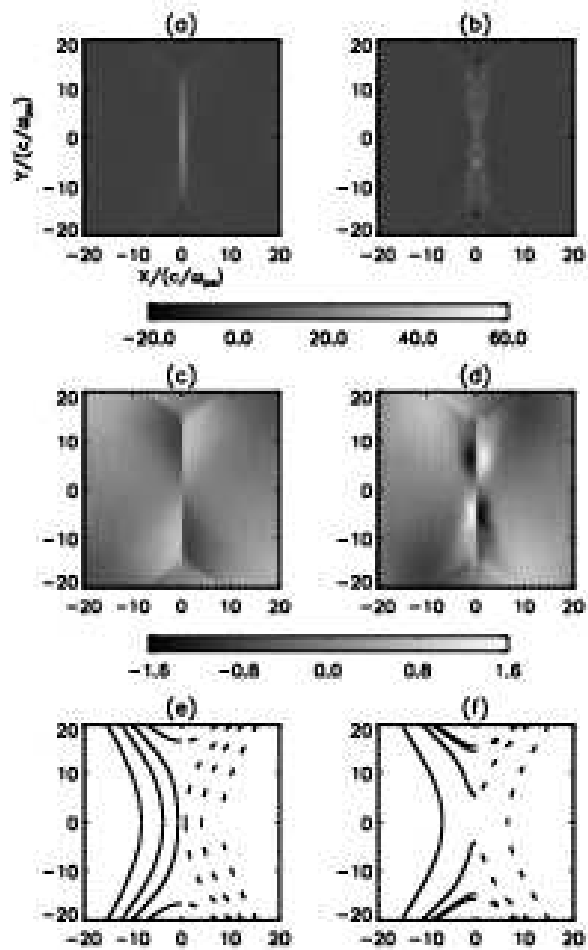}}
     \caption{ \label{Fig:jz_a224} Spatial distribution of total current density, 
             $j_z$, at (a) $t=45$ 
             and (b) $t=70$ for $\alpha = 2.24$.
             The total current density is normalised by the initial value, 
             $j_0 = n_0 e v_{d0}$.
             The same for out-of-plane magnetic field, 
             $B_z$, which is normalised to $B_0$ at 
	     (c) $t=45$ 
             and (d) $t=70$.
	     Panels (e) and (f) show individual magnetic field lines
             with different intensity, 
             $|B| / B_0 = 2.0$, $3.5$, $4.0$ and $4.5$ on the boundaries
	     at $t=45$ 
             and (d) $t=70$ (f) respectively.
             The left and right field lines in the simulation box 
             are distinguished 
             with solid and dotted line styles, respectively.}
    \end{figure} 
  
  Fig.~\ref{Fig:jz_a224} panels (a)  and (b) show spatial distribution of
  the $j_z$ current at the peak of the reconnection $t=45$,
  and after the current sheet decayed, $t=70$, [the first bounce -- see Fig.~\ref{Fig:timed_ezjz_a224}].
  The noticeable difference from panels (c) and (d) in Fig.~\ref{Fig:jz_a120}
  is that now current sheet becomes longer and thinner [the same conclusion reached in Ref.\cite{mcc96},
  in that large perturbations (strong compression) yields current sheet thinning and onset of
  more efficient reconnection]. As in the weakly stressed case, we also did the
  following calculation:
  Using data from Fig.~\ref{Fig:jz_a224}(a) (peak current time
  snapshot) the width and length of current sheet was estimated 
  by looking at  half-width of the $j_z(x,0)$ and $j_z(0,y)$  
  peak (in both $x$ and $y$-directions). The result is width,
  $\delta \simeq 0.4 $, and length, 
  $\Delta \simeq 17$
  The ratio of the two gives the inflow Mach number $M_A \simeq \delta / \Delta \simeq 0.02$.
  This is 125 times smaller that peak reconnection rate of 2.5 from Fig.~\ref{Fig:timed_ezjz_a224}(a)
  and 25 time smaller than average reconnection rate of 0.5 from the same figure.
  This discrepancy can only be attributed to the fact that formula 
  $M \simeq \delta / \Delta$ only applies to the steady reconnection process.
  In the case of $\alpha=1.2$ the reconnection process was relatively steady
  thus the latter formula proved a good match, but for $\alpha=2.24$ 
  the process is too dynamic and thus $M_A \simeq \delta / \Delta$ as a measure of reconnection fails.
  Fig.~\ref{Fig:jz_a224} panels (c)  and (d) again show
  familiar quadruple out-of-plane magnetic field structure, but now
  it is much more elongated. The panels (e) and (f) where we trace individual
  magnetic field lines at two different times provide another proof
  that the reconnection takes places and that the current sheet is now much
  longer and thinner.
    
  \begin{figure}
    \resizebox{\hsize}{!}{\includegraphics{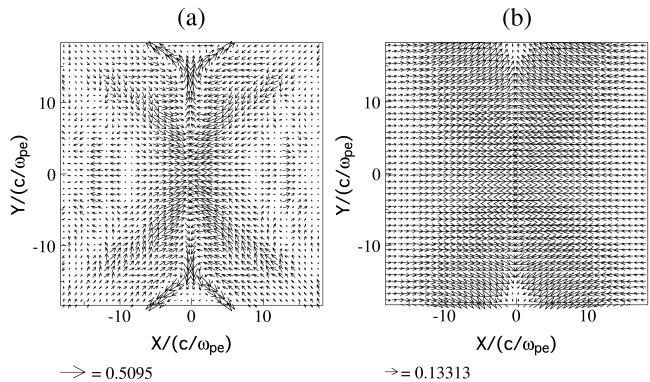}}
    \caption{ \label{Fig:fl_a224} The same as in Fig.~\ref{Fig:fl_a120} but for
    $\alpha=2.24$. Here $t=45$. Note that here $V_{A0} / c = 0.36$.}
     \end{figure}
     
 In Fig.~\ref{Fig:fl_a224} is an analog of Fig.~\ref{Fig:fl_a120}  but 
 for $\alpha=2.24$. 
  Noteworthy difference from the weak case is 
 that the plasma inflow into the current sheet is perpendicular to it, 
with the electron outflow speeds reaching   external  Alfv\'en  Mach number of $0.5/0.36=1.4$
[note arrow length in Fig.~\ref{Fig:fl_a224}(a)],
and ions again being about four times slower than electrons. 
Mind that now $V_{A0} / c = 0.36$.
     
  \begin{figure}
    \resizebox{\hsize}{!}{\includegraphics{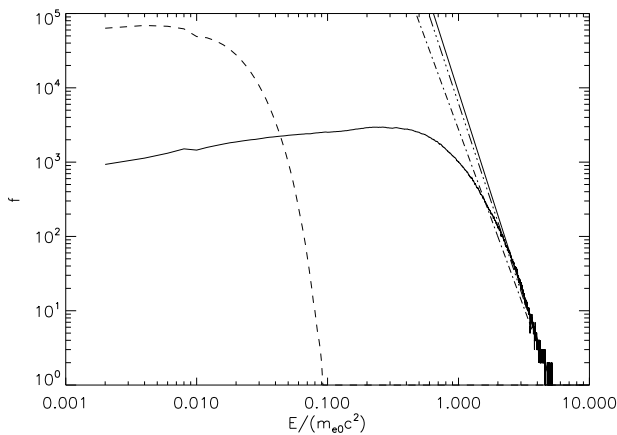}}
     \caption{\label{Fig:ed_a224} The same as in Fig.~\ref{Fig:ed_a120}
     but for $\alpha=2.24$. Here $t=250$.
     Data is produced using a region 
             near the maximum current generation 
             ($-1 \leq x \leq 1, 
               -16 \leq y \leq 16$).}
      \end{figure} 
  
  Fig.~\ref{Fig:ed_a224} shows the local electron energy distribution function 
  in the current sheet at $t = 0$ (dashed curve) 
  and $t = 250$ (solid curve) for $\alpha = 2.24$.
  Note that the dashed curves (corresponding to $t=0$) in Fig.~\ref{Fig:ed_a120} and
  Fig.~\ref{Fig:ed_a224} are different. This is due to the fact that different
  $\alpha$'s, mean different initial $j_0$ current [see Eq.(2)], which
  contribute to the calculation of the distribution function.  
  We performed numerical fit to the high energy part of the distribution function
  ($E > 2.4 m_{e0} c^2=1.226$ MeV) and found that the electron distribution has
  the power law index of -5.5 which is quite close to the observed power law range of 
  -4.8 and -5.3 \cite{oe02}. Note that now attained energies are much higher
 $4 \times m_{e0} c^2 \simeq 0.2 \times 511 \; {\rm keV}\simeq 2$ MeV
  (at the high end of the spectrum). Again, we note that values quoted in
  keV and MeV should be taken with caution [see discussion of Fig.~\ref{Fig:ed_a120} above].
   
\subsection{Energetics of the reconnection process}

\begin{figure}
    \resizebox{\hsize}{!}{\includegraphics{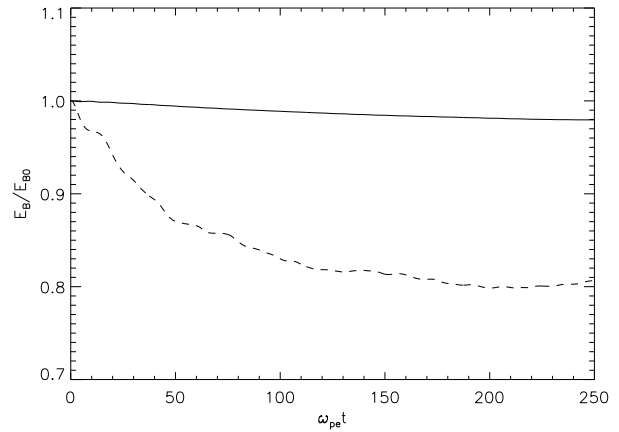}}
    \caption{ \label{Fig:e_mag} The magnetic energy,
$E_B (t)=\int \int (B_x2+B_y2+B_z2)/(2\mu_0) dxdy$, normalised on its initial value, $E_B (0)$, for two cases
$\alpha=1.2$ (solid line) and $\alpha=2.24$ (dashed line) versus time.
Note that $\omega_{pe}t=250$ corresponds to 1.25  Alfv\'en times).}
     \end{figure}

We now try to estimate the efficiently of the X-point collapse
by looking at the energetics of the process. 
In particular in Fig.~\ref{Fig:e_mag} we plot
the magnetic energy,
$E_B (t)=\int \int (B_x2+B_y2+B_z2)/(2\mu_0) dxdy$, for two cases
$\alpha=1.2$ (solid line) and $\alpha=2.24$ (dashed line).
The normalisation in each case is $E_B(0)$.
Note that the latter is different 
in the both cases because different stress means different initial magnetic field.
We gather from this graph that in the case of weak stress, 2\% of the initial
magnetic energy (which is a dominant part of the total energy because our plasma beta  is 0.02)
is converted into heat and energy of super-thermal (accelerated) particles. 
In the case of large perturbations (strong compression) 20 \%
in the initial magnetic energy is converted into heating and particle acceleration.

\section{Conclusion and Discussions}

We studied magnetic reconnection during collisionless, stressed,  X-point collapse 
using kinetic, 2.5D,  fully electromagnetic, relativistic Particle-in-Cell numerical code. 
We investigated two cases of weakly and strongly stressed  X-point.
Here by weakly and strongly we mean 20 \% and 124 \% unidirectional spatial 
compression of the X-point, respectively. The reconnection rate,  
defined as the out-of-plane electric field in the magnetic null normalised by 
the product of external magnetic field and Alfv\'en speeds,
peaks at 0.11 (at 0.85 Alfv\'en times), with its average over 1.25 Alfv\'en times being
0.04. We found that during the peak of the reconnection, electron inflow into the current sheet is mostly concentrated along
the separatrices until they deflect from the current sheet on the 
scale of electron skin depth, with the electron outflow speeds being of the order of 
the external Alfv\'en speed.
Ion inflow starts to deflect from the current sheet on the ion skin depth scale with
the outflow speeds about four times smaller than that of electrons.
Electron energy distribution in the current sheet, at the high energy end of the spectrum, 
shows a power law distribution with the index which varies in time, attaining a maximal value of -4.1 
at the final simulation time step (this corresponds to 1.25 Alfv\'en times). 

The obtained results in the strongly stressed case show that the 
magnetic reconnection peak occurs about 3.4 times faster (at 0.25 Alfv\'en times) and is more efficient.
The peak reconnection rate now
attains value 2.5 (at 0.25 Alfv\'en times), with the average reconnection rate over 1.25 Alfv\'en times being
0.5. Plasma inflow into the current sheet is perpendicular to it, 
with the electron outflow seeds reaching 1.4 Alfv\'en external Mach number 
and ions again being about four times slower than electrons. 
The power law energy spectrum for the 
electrons in the current sheet attains now a steeper index of -5.5. This is close to the typical
observed power law indexes in the vicinity of X-type region in the Earth's magneto-tail \cite{oe02}. 

The reconnection rate versus time figures in both cases indicate that the reconnection
has bursty, time-transient behaviour. 
In the considered two cases, 2\%  (in weakly stressed) and 20\% (in strongly stressed) 
of the initial magnetic energy is converted into heat and energy of accelerated particles,
respectively, both within about one Alfv\'en time.
This is somewhat similar to the previous resistive MHD analog \cite{mcc96} 
of the present study, in that
small perturbation (weak stress) initiates X-point collapse, but then
reconnection process is choked off by the gas pressure. At present, it is
unclear, however, whether we seem to observe similar behaviour in our simulation.
After all, e.g. Petschek mechanism reconnection can be choked off by
a different (other than pressure) 
mechanism [when the diffusion-region inflow magnetic field gets too small \cite{pf00}].

We also found that in the both cases, during the peak of the reconnection, 
the quadruple out-of-plane magnetic field 
is generated. This most is likely to suggest that Hall regime of the reconnection
takes place \cite{uk06}. 

In addition to the fundamental interest of converting magnetic energy into heat and
the energy of accelerated particles, these results are significant for 
e.g. solar coronal heating problem.
It has been estimated that even if only 2\% of the magnetic energy in the 
solar corona is converted into heat, the coronal heating problem would be
solved. Also, it is known that the typical resistive diffusion time in the
corona is $10^{15}$ Alfv\'en times ($10^8$ years), while the
time-transient phenomena such as flares (one of the possible candidates of
coronal heating) occur on time scales of 10 -- 100 Alfv\'en times.
Our results for the strongly stressed case suggest that 
20 \% of initial magnetic energy can be released in just 1  Alfv\'en time.
On one hand, one has to realise, that
the obtained results are for an X-point configuration, and naturally, 
the bulk of solar corona is not made of solely of
X-points. On the other hand, it is also known that heating  of the active regions
would provide circa 82 \% of the coronal heating budget \cite{asch07}.
In turn, given the fact that X-points are common occurrence in the active regions,
our results seem to be of importance for solving the coronal heating problem.

It important to stress that the obtained main result of the 20\% conversion of
the initial magnetic energy into heat and energy of super-thermal particles, within about one Alfv\'en time, 
is obtained in the collisionless
regime, thus this result does not suffer from any uncertainty in the anomalous resistivity 
as in the case of resistive MHD reconnection.

\acknowledgements
  Authors acknowledge use of computational facilities
  available through UKMHD consortium lead by University of St. Andrews.
  This research was funded by the 
  United Kingdom's Science and Technology Facilities Council.


\begin{thebibliography}{34}
\expandafter\ifx\csname natexlab\endcsname\relax\def\natexlab#1{#1}\fi
\expandafter\ifx\csname bibnamefont\endcsname\relax
  \def\bibnamefont#1{#1}\fi
\expandafter\ifx\csname bibfnamefont\endcsname\relax
  \def\bibfnamefont#1{#1}\fi
\expandafter\ifx\csname citenamefont\endcsname\relax
  \def\citenamefont#1{#1}\fi
\expandafter\ifx\csname url\endcsname\relax
  \def\url#1{\texttt{#1}}\fi
\expandafter\ifx\csname urlprefix\endcsname\relax\def\urlprefix{URL }\fi
\providecommand{\bibinfo}[2]{#2}
\providecommand{\eprint}[2][]{\url{#2}}

\bibitem[{\citenamefont{{Priest} and {Forbes}}(2000)}]{pf00}
\bibinfo{author}{\bibfnamefont{E.}~\bibnamefont{{Priest}}} \bibnamefont{and}
  \bibinfo{author}{\bibfnamefont{T.}~\bibnamefont{{Forbes}}},
  \emph{\bibinfo{title}{{Magnetic reconnection: MHD theory and applications}}}
  (\bibinfo{publisher}{Cambridge University Press}, \bibinfo{year}{2000}).

\bibitem[{\citenamefont{{Biskamp}}(2005)}]{b05}
\bibinfo{author}{\bibfnamefont{D.}~\bibnamefont{{Biskamp}}},
  \emph{\bibinfo{title}{{Magnetic reconnection in Plasmas}}}
  (\bibinfo{publisher}{Cambridge University Press}, \bibinfo{year}{2005}).

\bibitem[{\citenamefont{{Birn} and {Priest}}(2007)}]{bp07}
\bibinfo{author}{\bibfnamefont{J.}~\bibnamefont{{Birn}}} \bibnamefont{and}
  \bibinfo{author}{\bibfnamefont{E.~R.} \bibnamefont{{Priest}}},
  \emph{\bibinfo{title}{{Reconnection of magnetic fields: magnetohydrodynamics
  and collisionless theory and observations}}} (\bibinfo{publisher}{Cambridge:
  Cambridge University Press}, \bibinfo{year}{2007}).

\bibitem[{\citenamefont{{Birn} et~al.}(2001)\citenamefont{{Birn}, {Drake},
  {Shay}, and et~al.}}]{birn2001}
\bibinfo{author}{\bibfnamefont{J.}~\bibnamefont{{Birn}}},
  \bibinfo{author}{\bibfnamefont{J.~F.} \bibnamefont{{Drake}}},
  \bibinfo{author}{\bibfnamefont{M.~A.} \bibnamefont{{Shay}}},
  \bibnamefont{and} \bibinfo{author}{\bibnamefont{et~al.}},
  \bibinfo{journal}{J. \ Geophys. \ Res.} \textbf{\bibinfo{volume}{106}},
  \bibinfo{pages}{3715} (\bibinfo{year}{2001}).

\bibitem[{\citenamefont{{Pritchett}}(2001)}]{pritchett2001}
\bibinfo{author}{\bibfnamefont{P.~L.} \bibnamefont{{Pritchett}}},
  \bibinfo{journal}{J. \ Geophys. \ Res.} \textbf{\bibinfo{volume}{106}},
  \bibinfo{pages}{3783} (\bibinfo{year}{2001}).

\bibitem[{\citenamefont{{Birn} et~al.}(2005)\citenamefont{{Birn}, {Galsgaard},
  {Hesse}, and et~al.}}]{birn2005}
\bibinfo{author}{\bibfnamefont{J.}~\bibnamefont{{Birn}}},
  \bibinfo{author}{\bibfnamefont{K.}~\bibnamefont{{Galsgaard}}},
  \bibinfo{author}{\bibfnamefont{M.}~\bibnamefont{{Hesse}}}, \bibnamefont{and}
  \bibinfo{author}{\bibnamefont{et~al.}}, \bibinfo{journal}{Geophys. \ Res. \
  Lett.} \textbf{\bibinfo{volume}{32}}, \bibinfo{pages}{L06105}
  (\bibinfo{year}{2005}).

\bibitem[{\citenamefont{{Arber} and {Haynes}}(2006)}]{ah06}
\bibinfo{author}{\bibfnamefont{T.~D.} \bibnamefont{{Arber}}} \bibnamefont{and}
  \bibinfo{author}{\bibfnamefont{M.}~\bibnamefont{{Haynes}}},
  \bibinfo{journal}{Physics of Plasmas} \textbf{\bibinfo{volume}{13}},
  \bibinfo{pages}{2105} (\bibinfo{year}{2006}).

\bibitem[{\citenamefont{{McClements} et~al.}(2006)\citenamefont{{McClements},
  {Shah}, and {Thyagaraja}}}]{ken06}
\bibinfo{author}{\bibfnamefont{K.~G.} \bibnamefont{{McClements}}},
  \bibinfo{author}{\bibfnamefont{N.}~\bibnamefont{{Shah}}}, \bibnamefont{and}
  \bibinfo{author}{\bibfnamefont{A.}~\bibnamefont{{Thyagaraja}}},
  \bibinfo{journal}{J. \ Plasma \ Phys.} \textbf{\bibinfo{volume}{72}},
  \bibinfo{pages}{571} (\bibinfo{year}{2006}).

\bibitem[{\citenamefont{{Vekstein} and {Browning}}(1997)}]{vb97}
\bibinfo{author}{\bibfnamefont{G.~E.} \bibnamefont{{Vekstein}}}
  \bibnamefont{and} \bibinfo{author}{\bibfnamefont{P.~K.}
  \bibnamefont{{Browning}}}, \bibinfo{journal}{Physics of Plasmas}
  \textbf{\bibinfo{volume}{4}}, \bibinfo{pages}{2261} (\bibinfo{year}{1997}).

\bibitem[{\citenamefont{{Browning} and {Vekstein}}(2001)}]{bv01}
\bibinfo{author}{\bibfnamefont{P.~K.} \bibnamefont{{Browning}}}
  \bibnamefont{and} \bibinfo{author}{\bibfnamefont{G.~E.}
  \bibnamefont{{Vekstein}}}, \bibinfo{journal}{J. \ Geophys. \ Res.}
  \textbf{\bibinfo{volume}{106}}, \bibinfo{pages}{18677}
  (\bibinfo{year}{2001}).

\bibitem[{\citenamefont{{Zharkova} and {Gordovskyy}}(2004)}]{zhar04}
\bibinfo{author}{\bibfnamefont{V.~V.} \bibnamefont{{Zharkova}}}
  \bibnamefont{and}
  \bibinfo{author}{\bibfnamefont{M.}~\bibnamefont{{Gordovskyy}}},
  \bibinfo{journal}{\apj} \textbf{\bibinfo{volume}{604}}, \bibinfo{pages}{884}
  (\bibinfo{year}{2004}).

\bibitem[{\citenamefont{{Hamilton} et~al.}(2005)\citenamefont{{Hamilton},
  {Fletcher}, {McClements}, and et~al.}}]{ham05}
\bibinfo{author}{\bibfnamefont{B.}~\bibnamefont{{Hamilton}}},
  \bibinfo{author}{\bibfnamefont{L.}~\bibnamefont{{Fletcher}}},
  \bibinfo{author}{\bibfnamefont{K.~G.} \bibnamefont{{McClements}}},
  \bibnamefont{and} \bibinfo{author}{\bibnamefont{et~al.}},
  \bibinfo{journal}{\apj} \textbf{\bibinfo{volume}{625}}, \bibinfo{pages}{496}
  (\bibinfo{year}{2005}).

\bibitem[{\citenamefont{{Dalla} and {Browning}}(2005)}]{db05}
\bibinfo{author}{\bibfnamefont{S.}~\bibnamefont{{Dalla}}} \bibnamefont{and}
  \bibinfo{author}{\bibfnamefont{P.~K.} \bibnamefont{{Browning}}},
  \bibinfo{journal}{Astron. \ Astrophys.} \textbf{\bibinfo{volume}{436}},
  \bibinfo{pages}{1103} (\bibinfo{year}{2005}).

\bibitem[{\citenamefont{{Dalla} and {Browning}}(2006)}]{db06}
\bibinfo{author}{\bibfnamefont{S.}~\bibnamefont{{Dalla}}} \bibnamefont{and}
  \bibinfo{author}{\bibfnamefont{P.~K.} \bibnamefont{{Browning}}},
  \bibinfo{journal}{Astrophys. \ J. \ Lett.} \textbf{\bibinfo{volume}{640}},
  \bibinfo{pages}{L99} (\bibinfo{year}{2006}).

\bibitem[{\citenamefont{{Browning} and {Dalla}}(2007)}]{bd07}
\bibinfo{author}{\bibfnamefont{P.}~\bibnamefont{{Browning}}} \bibnamefont{and}
  \bibinfo{author}{\bibfnamefont{S.}~\bibnamefont{{Dalla}}},
  \bibinfo{journal}{Mem. \ Soc. \ Astron. \ Italiana}
  \textbf{\bibinfo{volume}{78}}, \bibinfo{pages}{255} (\bibinfo{year}{2007}).

\bibitem[{\citenamefont{{Litvinenko}}(2006)}]{lit06}
\bibinfo{author}{\bibfnamefont{Y.~E.} \bibnamefont{{Litvinenko}}},
  \bibinfo{journal}{Astron. \ Astrophys.} \textbf{\bibinfo{volume}{452}},
  \bibinfo{pages}{1069} (\bibinfo{year}{2006}).

\bibitem[{\citenamefont{{Dungey}}(1953)}]{d53}
\bibinfo{author}{\bibfnamefont{J.~W.} \bibnamefont{{Dungey}}},
  \bibinfo{journal}{Phil. \ Mag.} \textbf{\bibinfo{volume}{44}},
  \bibinfo{pages}{725} (\bibinfo{year}{1953}).

\bibitem[{\citenamefont{{Semenov} et~al.}(1983)\citenamefont{{Semenov}, {Heyn},
  and {Kubyshkin}}}]{s83}
\bibinfo{author}{\bibfnamefont{V.~S.} \bibnamefont{{Semenov}}},
  \bibinfo{author}{\bibfnamefont{M.~F.} \bibnamefont{{Heyn}}},
  \bibnamefont{and} \bibinfo{author}{\bibfnamefont{I.~V.}
  \bibnamefont{{Kubyshkin}}}, \bibinfo{journal}{Sov. \ Astron.}
  \textbf{\bibinfo{volume}{27}}, \bibinfo{pages}{660} (\bibinfo{year}{1983}).

\bibitem[{\citenamefont{{Vekstein} and {Bian}}(2005)}]{vb05}
\bibinfo{author}{\bibfnamefont{G.}~\bibnamefont{{Vekstein}}} \bibnamefont{and}
  \bibinfo{author}{\bibfnamefont{N.}~\bibnamefont{{Bian}}},
  \bibinfo{journal}{\apj} \textbf{\bibinfo{volume}{632}}, \bibinfo{pages}{L151}
  (\bibinfo{year}{2005}).

\bibitem[{\citenamefont{{Ofman} et~al.}(1993)\citenamefont{{Ofman}, {Morrison},
  and {Steinolfson}}}]{oms93}
\bibinfo{author}{\bibfnamefont{L.}~\bibnamefont{{Ofman}}},
  \bibinfo{author}{\bibfnamefont{P.~J.} \bibnamefont{{Morrison}}},
  \bibnamefont{and} \bibinfo{author}{\bibfnamefont{R.~S.}
  \bibnamefont{{Steinolfson}}}, \bibinfo{journal}{\apj}
  \textbf{\bibinfo{volume}{417}}, \bibinfo{pages}{748} (\bibinfo{year}{1993}).

\bibitem[{\citenamefont{{McClymont} and {Craig}}(1996)}]{mcc96}
\bibinfo{author}{\bibfnamefont{A.~N.} \bibnamefont{{McClymont}}}
  \bibnamefont{and} \bibinfo{author}{\bibfnamefont{I.~J.~D.}
  \bibnamefont{{Craig}}}, \bibinfo{journal}{\apj}
  \textbf{\bibinfo{volume}{466}}, \bibinfo{pages}{487} (\bibinfo{year}{1996}).

\bibitem[{\citenamefont{{Forbes} and {Priest}}(1995)}]{fp95}
\bibinfo{author}{\bibfnamefont{T.~G.} \bibnamefont{{Forbes}}} \bibnamefont{and}
  \bibinfo{author}{\bibfnamefont{E.~R.} \bibnamefont{{Priest}}},
  \bibinfo{journal}{\apj} \textbf{\bibinfo{volume}{446}}, \bibinfo{pages}{377}
  (\bibinfo{year}{1995}).

\bibitem[{\citenamefont{{Sakai} and {Tanaka}}(2007)}]{sakai_tanaka2007}
\bibinfo{author}{\bibfnamefont{J.~I.} \bibnamefont{{Sakai}}} \bibnamefont{and}
  \bibinfo{author}{\bibfnamefont{Y.}~\bibnamefont{{Tanaka}}},
  \bibinfo{journal}{Astron. \ Astrophys.} \textbf{\bibinfo{volume}{468}},
  \bibinfo{pages}{1075} (\bibinfo{year}{2007}).

\bibitem[{\citenamefont{{Buneman}}(1993)}]{buneman1993}
\bibinfo{author}{\bibfnamefont{O.}~\bibnamefont{{Buneman}}},
  \emph{\bibinfo{title}{{Computer Space Plasma Physics, Simulation Techniques
  and Software ed. H. Matsumoto \& Y. Omura}}} (\bibinfo{publisher}{Terra
  Scientific}, \bibinfo{year}{1993}).

\bibitem[{\citenamefont{Tsiklauri
  et~al.}(2005{\natexlab{a}})\citenamefont{Tsiklauri, Sakai, and
  Saito}}]{tss05a}
\bibinfo{author}{\bibfnamefont{D.}~\bibnamefont{Tsiklauri}},
  \bibinfo{author}{\bibfnamefont{J.~I.} \bibnamefont{Sakai}}, \bibnamefont{and}
  \bibinfo{author}{\bibfnamefont{S.}~\bibnamefont{Saito}},
  \bibinfo{journal}{New\ J.\ Phys.} \textbf{\bibinfo{volume}{7}},
  \bibinfo{pages}{79} (\bibinfo{year}{2005}{\natexlab{a}}).

\bibitem[{\citenamefont{Tsiklauri
  et~al.}(2005{\natexlab{b}})\citenamefont{Tsiklauri, Sakai, and
  Saito}}]{tss05b}
\bibinfo{author}{\bibfnamefont{D.}~\bibnamefont{Tsiklauri}},
  \bibinfo{author}{\bibfnamefont{J.~I.} \bibnamefont{Sakai}}, \bibnamefont{and}
  \bibinfo{author}{\bibfnamefont{S.}~\bibnamefont{Saito}},
  \bibinfo{journal}{Astron.\ Astrophys.} \textbf{\bibinfo{volume}{435}},
  \bibinfo{pages}{1105} (\bibinfo{year}{2005}{\natexlab{b}}).

\bibitem[{\citenamefont{{Heyvaerts} and {Priest}}(1983)}]{hp83}
\bibinfo{author}{\bibfnamefont{J.}~\bibnamefont{{Heyvaerts}}} \bibnamefont{and}
  \bibinfo{author}{\bibfnamefont{E.~R.} \bibnamefont{{Priest}}},
  \bibinfo{journal}{Astron. \ Astrophys.} \textbf{\bibinfo{volume}{117}},
  \bibinfo{pages}{220} (\bibinfo{year}{1983}).

\bibitem[{\citenamefont{{Uzdensky} and {Kulsrud}}(2006)}]{uk06}
\bibinfo{author}{\bibfnamefont{D.~A.} \bibnamefont{{Uzdensky}}}
  \bibnamefont{and} \bibinfo{author}{\bibfnamefont{R.~M.}
  \bibnamefont{{Kulsrud}}}, \bibinfo{journal}{Physics of Plasmas}
  \textbf{\bibinfo{volume}{13}}, \bibinfo{pages}{2305} (\bibinfo{year}{2006}).

\bibitem[{\citenamefont{{Shay} et~al.}(2001)\citenamefont{{Shay}, {Drake},
  {Rogers}, and et~al.}}]{shay01}
\bibinfo{author}{\bibfnamefont{M.~A.} \bibnamefont{{Shay}}},
  \bibinfo{author}{\bibfnamefont{J.~F.} \bibnamefont{{Drake}}},
  \bibinfo{author}{\bibfnamefont{B.~N.} \bibnamefont{{Rogers}}},
  \bibnamefont{and} \bibinfo{author}{\bibnamefont{et~al.}},
  \bibinfo{journal}{J. \ Geophys. \ Res.} \textbf{\bibinfo{volume}{106}},
  \bibinfo{pages}{3759} (\bibinfo{year}{2001}).

\bibitem[{\citenamefont{\O{}ieroset et~al.}(2002)\citenamefont{\O{}ieroset,
  Lin, Phan, and et~al.}}]{oe02}
\bibinfo{author}{\bibfnamefont{M.}~\bibnamefont{\O{}ieroset}},
  \bibinfo{author}{\bibfnamefont{R.~P.} \bibnamefont{Lin}},
  \bibinfo{author}{\bibfnamefont{T.~D.} \bibnamefont{Phan}}, \bibnamefont{and}
  \bibinfo{author}{\bibnamefont{et~al.}}, \bibinfo{journal}{Phys. Rev. Lett.}
  \textbf{\bibinfo{volume}{89}}, \bibinfo{pages}{195001}
  (\bibinfo{year}{2002}).

\bibitem[{\citenamefont{{Ma} and {Summers}}(1998)}]{ms98}
\bibinfo{author}{\bibfnamefont{C.-Y.} \bibnamefont{{Ma}}} \bibnamefont{and}
  \bibinfo{author}{\bibfnamefont{D.}~\bibnamefont{{Summers}}},
  \bibinfo{journal}{Geophys. \ Res. \ Lett.} \textbf{\bibinfo{volume}{25}},
  \bibinfo{pages}{4099} (\bibinfo{year}{1998}).

\bibitem[{\citenamefont{{Nakariakov} et~al.}(2006)\citenamefont{{Nakariakov},
  {Foullon}, {Verwichte}, and et~al.}}]{nakariakov2006}
\bibinfo{author}{\bibfnamefont{V.~M.} \bibnamefont{{Nakariakov}}},
  \bibinfo{author}{\bibfnamefont{C.}~\bibnamefont{{Foullon}}},
  \bibinfo{author}{\bibfnamefont{E.}~\bibnamefont{{Verwichte}}},
  \bibnamefont{and} \bibinfo{author}{\bibnamefont{et~al.}},
  \bibinfo{journal}{Astron. \ Astrophys.} \textbf{\bibinfo{volume}{452}},
  \bibinfo{pages}{343} (\bibinfo{year}{2006}).

\bibitem[{\citenamefont{{Ofman} and {Sui}}(2006)}]{ofman_sui2006}
\bibinfo{author}{\bibfnamefont{L.}~\bibnamefont{{Ofman}}} \bibnamefont{and}
  \bibinfo{author}{\bibfnamefont{L.}~\bibnamefont{{Sui}}},
  \bibinfo{journal}{\apj} \textbf{\bibinfo{volume}{644}}, \bibinfo{pages}{L149}
  (\bibinfo{year}{2006}).

\bibitem[{\citenamefont{{Aschwanden} et~al.}(2007)\citenamefont{{Aschwanden},
  {Winebarger}, {Tsiklauri}, and et~al.}}]{asch07}
\bibinfo{author}{\bibfnamefont{M.~J.} \bibnamefont{{Aschwanden}}},
  \bibinfo{author}{\bibfnamefont{A.}~\bibnamefont{{Winebarger}}},
  \bibinfo{author}{\bibfnamefont{D.}~\bibnamefont{{Tsiklauri}}},
  \bibnamefont{and} \bibinfo{author}{\bibnamefont{et~al.}},
  \bibinfo{journal}{\apj} \textbf{\bibinfo{volume}{659}}, \bibinfo{pages}{1673}
  (\bibinfo{year}{2007}).

\end{thebibliography}
\end{document}